%% file: main.tex
\documentclass[acmsmall]{acmart}
\usepackage[suppress]{color-edits} 

\addauthor{km}{blue} 
\addauthor{si}{purple} 
\addauthor{eh}{red} 

\AtBeginDocument{%
  \providecommand\BibTeX{{%
    \normalfont B\kern-0.5em{\scshape i\kern-0.25em b}\kern-0.8em\TeX}}}

\setcopyright{acmcopyright}
\copyrightyear{2023}
\acmYear{2023}
\acmDOI{XXXXXXX.XXXXXXX}

\acmConference[CSCW '24]{ACM Computer-Supported Collaborative Work}{November 9--13,
  2024}{San José, Costa Rica}
\acmPrice{15.00}
\acmISBN{978-1-4503-XXXX-X/18/06}

\begin{document}





\title[AI-Powered Reminders for Collaborative Tasks]{AI-Powered Reminders for Collaborative Tasks:\\ Experiences and Futures}


\author{Katelyn Morrison}
\orcid{0000-0002-2644-4422}
\email{kcmorris@cs.cmu.edu}
\affiliation{%
  \institution{Carnegie Mellon University}
  \city{Pittsburgh}
  \state{Pennsylvania}
  \country{USA}
}

\author{Shamsi Iqbal}
\orcid{0000-0002-9380-8759}
\email{shamsi@microsoft.com}

\author{Eric Horvitz}
\orcid{0000-0002-8823-0614}
\affiliation{%
  \email{horvitz@microsoft.com}
   \institution{Microsoft}
  \city{Redmond}
  \state{Washington}
  \country{USA}
}

\renewcommand{\shortauthors}{Morrison, et al.}

\begin{abstract}
  Email continues to serve as a central medium for managing collaborations. While unstructured email messaging is lightweight and conducive to coordination, it is easy to overlook commitments and requests for collaborations that are embedded in the text of free-flowing communications. Twenty-one years ago,~\citet{bellotti_taskmaster} proposed TaskMaster with the goal of redesigning the email interface to have explicit task management capabilities. Recently, AI-based task recognition and reminder services have been introduced in major email systems as one approach to managing asynchronous collaborations. While these services have been provided to millions of people around the world, there is little understanding of how people interact with and benefit from them. We explore knowledge workers' experiences with Microsoft's Viva Daily Briefing Email to better understand how AI-powered reminders can support asynchronous collaborations. Through semi-structured interviews and surveys, we shed light on how AI-powered reminders are incorporated into workflows to support asynchronous collaborations. We identify what knowledge workers prefer AI-powered reminders to remind them about and how they would like to interact with these reminders. Using mixed methods and a self-assessment methodology, we investigate the relationship between information workers’ work styles and the perceived value of the Viva Daily Briefing Email to identify users who are more likely to benefit from AI-powered reminders for asynchronous collaborations. We conclude by discussing the experiences and futures of AI-powered reminders for collaborative tasks and asynchronous collaborations.
\end{abstract}



\begin{teaserfigure}
    \centering
  \includegraphics[trim={0cm 7cm 12cm 0cm},width=\textwidth,clip]{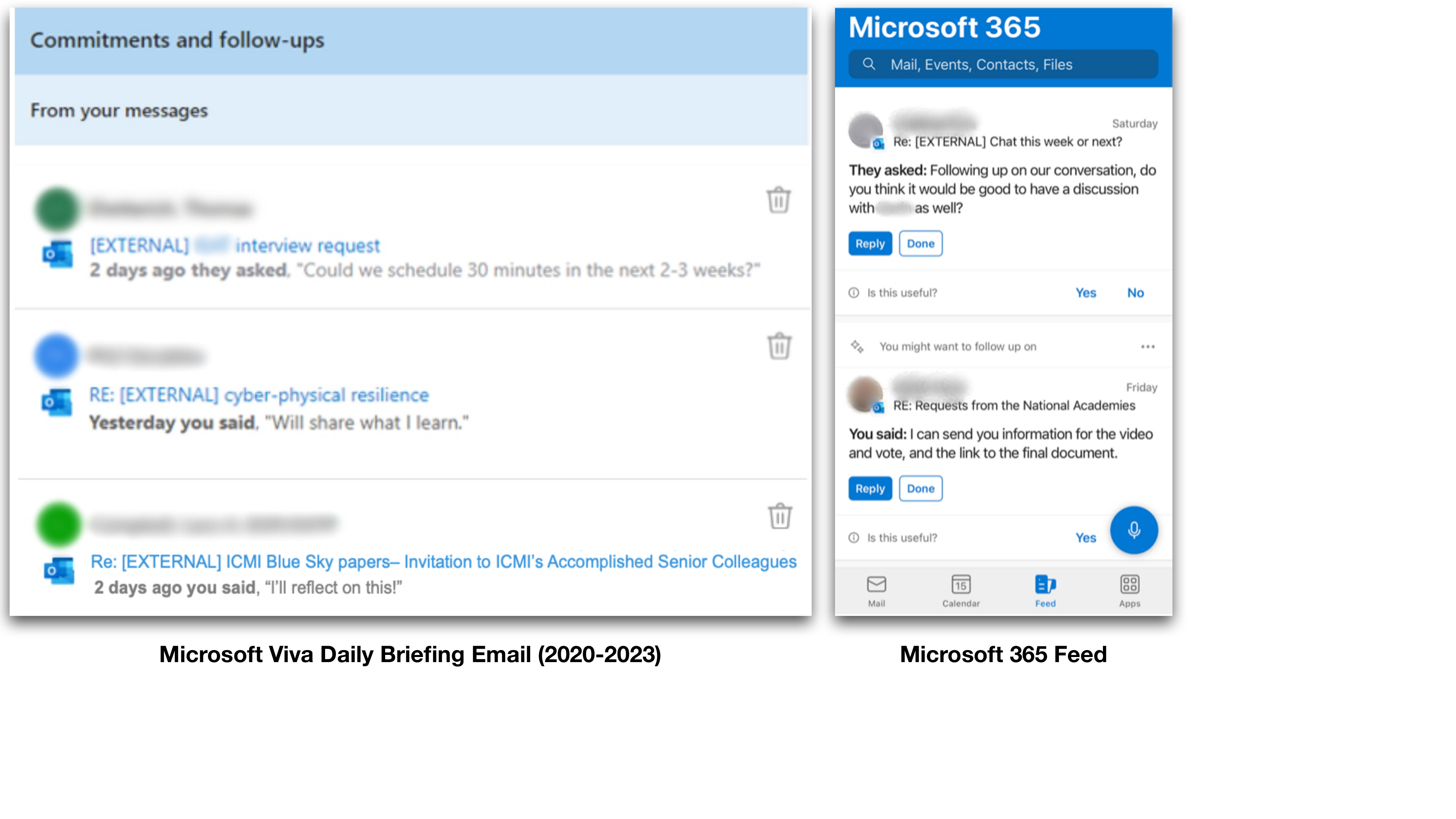}
  \caption{Left: Reminders about commitments and requests surfaced in a daily email digest by a service that employs machine learning to provide users with reminders about commitments and requests drawn from their unstructured email communications. Right: Recent update showing reminders about commitments and requests displayed in an always-available feed. Reminders displayed are drawn from E.H.'s email. 
  }
  \label{fig:teaser}
\end{teaserfigure}

\maketitle

\input{01-introduction}
\input{02-related-works}

\input{03-experiences}
\input{04-work-styles}

\input{05-interactions}

\input{06-discussion}
\input{07-limitations}
\input{08-conclusion}

\bibliographystyle{ACM-Reference-Format}
\bibliography{sample-base}

\end{document}

%% file: 01-introduction.tex
\section{Introduction}

Asynchronous collaborations are increasingly present in the modern workplace. Although asynchronous collaborations have taken many different forms, email is the most common. A recent study showed that the number of emails increased significantly following the start of the COVID-19 pandemic~\cite{DeFilippis_Impink_Singell_Polzer_Sadun_2022}.  More than ever before, email applications serve as central hubs for asynchronous communication, with numerous messages containing scattered information about collaborative efforts~\cite{dabbish2006email,ducheneaut2001mail,mackay1988more,bellotti2005quality}. In particular, inboxes and outboxes can be viewed as ill-structured repositories of commitments and requests. In busy environments, commitments made through the informality of natural language can easily slip through the cracks. While delays and ``dropping the ball'' are often understandable, forgetting to follow through on a commitment or request expressed in an email can slow progress on projects and deteriorate collaborations. Failures to address tasks in a timely manner can also lead to challenges with reputation and trust, as inappropriate delays and missed deadlines signal unreliability and disinterest~\cite{10.3389/fpsyg.2014.00657}. The dual of increasing reliance on email for task specification, communication, and management and the increasing volume of email motivates the potential value of AI-powered tools that help people identify, organize, and address collaborative tasks that are referred to in email communications.

Information workers have relied on memory augmentation tools such as physical and digital sticky notes, to-do list applications, and calendars to manage their requests and commitments\footnote{We will use the words `commitments' and `tasks' interchangeably throughout since commitments in the Viva Daily Briefing Email make up tasks.} made in emails throughout the day~\cite{10.1145/985692.985785}. However, these tools are only useful to someone if they remember to write down the commitment, such as commitments asserted in the natural language of email communications. Unfortunately, not all important commitments get added to a physical or digital to-do list. Interruptions have been cited as a reason why people in workplace settings may forget to do a task or add it to their to-do list~\cite{dismukes2012prospective}. Ultimately, commitments that do not make it to a physical or digital to-do list have a higher chance of ``falling through the cracks'' or being forgotten. As a result, these physical and digital reminder systems can only provide limited assistance regarding tasks discussed and managed via email or instant messaging platforms.

Reminder systems in the form of alerts provided by virtual assistants fielded by Apple, Google, Amazon, and Microsoft have been deployed widely \cite{manseau2020ai}. These services can complement or replace physical and digital to-do lists by helping users identify and remember tasks or deadlines. 
We study Microsoft's Viva Daily Briefing Email (Viva Briefing), which has been fielded within Microsoft Outlook, focusing on the email's ``Commitments and follow-ups'' section. The daily briefing's ``Commitments and follow-ups'' section is generated via natural language analysis and machine learning applied to extracting commitments and requests specified in natural language from email conversations. The goal is to remind users by providing a list of these automatically identified tasks in a recurring daily briefing email. The Viva Briefing was widely available to Microsoft Outlook users between 2020 and 2023 and now appears in the Microsoft 365 Feed. Despite the wide availability, no research has been published on the uses and influences of the service to support asynchronous collaborations in the workplace. 

As an illustrative example of how the Viva Daily Briefing Email can be used in the workplace, one of the authors (E.H.) received a reminder from his Viva Daily Briefing Email about an invitation from a conference program chair to submit his research to a special track at a conference (shown on the bottom left in Figure \ref{fig:teaser}).
The Viva Briefing reminded him that he had said he would reflect on the invitation, but once he replied to the email, the task ``fell through the cracks'' as a deadline loomed because it was not added to a physical or digital to-do list. This task was not considered an acutely important task at the time of the initial email. However, by overlooking the task amidst a busy schedule, he would have missed out on an opportunity to share unpublished findings with the community. Luckily, the Viva Briefing reminded him about his ``commitment'' to reflecting on and responding to submitting something to the conference. He attributes a timely reminder from the Viva Briefing as being the key factor leading to the allocation of time for paper drafting and submission under the deadline. The paper went on to receive an award, and the ideas were featured widely in the technical press~\cite{horvitz2022horizon}. The Viva Daily Briefing Email has also proven to be essential to asynchronous collaborative tasks. For example, one knowledge worker that we interviewed described to us how they rely on the reminders that are surfaced in the Viva Daily Briefing Email every day to complete commitments they made to their teammates.  

These examples highlight the value of the core functionality captured in the Viva Briefing and foreshadow directions for achieving more powerful and effective task management services in the future. Furthermore, fast-paced advances in AI, particularly with generative neural models, are framing more powerful directions for supporting people with commitments and collaborations. We need to pursue a deeper understanding of human-AI interaction workflows in support of asynchronous collaborations.

 The rise of asynchronous collaboration in the workplace has demonstrated the need to understand the usability challenges in task management tools and challenges researchers to understand what makes AI-powered reminders valuable to asynchronous collaborations. Furthermore, Khaokaew et al. point out that understanding what people desire from task management tools is underexplored~\cite{khaokaew2022imagining}. Tangentially, the rapid advancements with large language models motivate the need to comprehensively understand their impact on task management in the workplace. With that, our research aims to answer four research questions: 
 \begin{itemize}
     \item (\textbf{RQ1}) How have knowledge workers incorporated AI-powered reminders in their workflow to improve their asynchronous collaborations?
     \item (\textbf{RQ2}) What type of information do knowledge workers want AI-powered reminders to remind them about?
     \item (\textbf{RQ3}) How does an individual's workstyle impact how valuable AI-powered reminders are to them in the workplace?
     \item (\textbf{RQ4}) How do knowledge workers want to interact with AI-powered reminders to support their asynchronous collaborations?
 \end{itemize}   

\begin{figure}[!b]
    \includegraphics[trim=0cm 3cm 0cm 0cm, width=\linewidth, clip]{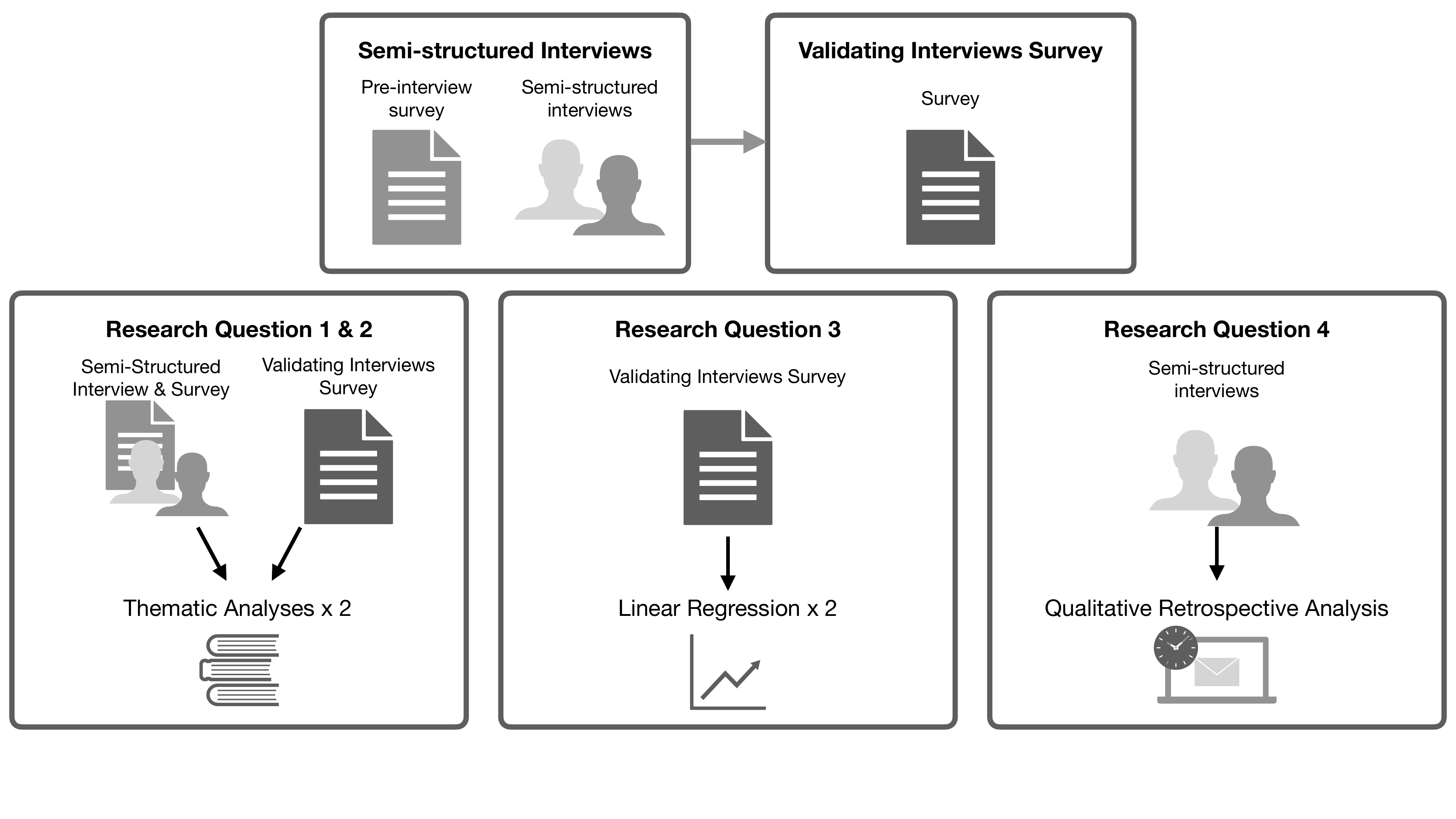}
    \caption{Summary of the methods and analyses for each research question. We answer the first two research questions by conducting two thematic analyses based on open-ended responses from the pre-interview and validating interview surveys. We create two linear regression models based on the validating interview survey questions to answer the third research question. We qualitatively analyze the retrospective portion of the semi-structured interviews to answer research question four. All studies were approved by an ethics and privacy review board (IRB).}
    \label{methods}
    \Description{}
\end{figure}

We focus on information workers at Microsoft 
who have the Viva Daily Briefing Email feature enabled in Microsoft Outlook for all of its employees. We chose to look at knowledge workers specifically because we believe they tend to work on a variety of smaller and larger projects at the same time that require asynchronous collaborations, resulting in many tasks and requests sent through email. We conducted semi-structured interviews and surveys to answer our research questions. First, we conducted 11 semi-structured interviews with a pre-interview survey to gain a deeper understanding of how AI-powered reminders have been incorporated into workflows to support asynchronous collaborations. These interviews were also used to understand if the tool ever surfaced something that was forgotten about to understand what people would prefer to be reminded about. These interviews also provided insight into how knowledge workers would like to interact with AI-powered reminders in the workplace. After these interviews, we sent out a survey to a broader population of information workers at Microsoft to validate our interview findings and gather more granular data\footnote{This survey is referred to as the validating interviews survey throughout this paper.}. Through a combination of the open-ended responses from the pre-interview survey and 45 responses from open-ended questions in the validating interviews survey, 
we address \textbf{RQ1} by presenting a thematic analysis of how the Viva Daily Briefing Email has been used in the workplace. We also highlight workflows mentioned by our interview participants. We address \textbf{RQ2} with a second thematic analysis of open-ended responses from both surveys about the type of information that users prefer to be reminded about. We perform a quantitative analysis on self-assessment data from the validating interviews survey to answer \textbf{RQ3}. Lastly, we capture how knowledge workers would like to interact with these AI-powered reminders through retrospective discussions about previously surfaced reminders during the semi-structured interviews to answer \textbf{RQ4}. A summary of the methods used to answer each research question is depicted in Figure \ref{methods}.


\textbf{Contributions.} In summary, we make several contributions to computer-supported cooperative work (CSCW). First, we identify how knowledge workers incorporated AI-powered reminders into their daily workflow, and expose how the reminders have impacted their asynchronous collaborations. Second, we identify the types of information that knowledge workers prefer AI-powered reminders to remind them about. Third, we identify how work styles can influence the perceived value and benefit of a daily briefing email that contains reminders on inferred commitments. Lastly, we share insights gleaned from our interviews about how knowledge workers would like to interact with  information surfaced by AI-powered reminders. We probe how AI-powered reminders can support asynchronous collaborations in the workplace among people with different work styles and suggest informal guidelines to consider when designing AI-powered reminders for collaborative work. 

%% file: 02-related-works.tex
\section{Evolution of Task Management in the Workplace}

Task management tools can support effective collaborations in the workplace. They provide a means for people to save commitments from team meetings and one-on-one conversations. While people have used computer-based tools to support asynchronous collaborations for many years, the features and capabilities of these tools have changed significantly. We briefly introduce prospective memory augmentation and its importance to asynchronous collaborations in the workplace. Then, we provide a brief history of task management tools that intend to support asynchronous collaborations and how they have changed over time. Lastly, we refer to other studies that have evaluated how people employ intelligent task management tools to support asynchronous collaborations in the workplace over the years.

\subsection{Importance of Prospective Memory Augmentation}

 The performance of an individual's \emph{prospective memory}, or the ability to remember to do something at an appropriate time in the future, is core to going about the work day~\cite{dismukes2012prospective, eldridge1992memory}. Shimamura et al.~\cite{Shimamura1991WhatIT} posit that prospective memory impacts planning and decision-making. Several lab-based studies have pursued understanding when our prospective memory fails and why, but little research has been conducted in the workplace  \cite{dismukes2012prospective}. Receiving a high volume of emails has been associated with email overload and stress~\cite{mcmurtry2014managing}, and stress has been shown to negatively affect one's ability to remember to do certain tasks~\cite{ihle2012age, chen2019effect}. Several works have studied the impact that interruptions and reminders in the workplace have on task management and productivity~\cite{dismukes2012prospective,horvitz2001notification,iqbal2007disruption,iqbal2010notifications,paul2015interruptive}. Along this line, one lab-based experiment demonstrated decreases in prospective memory performance as subjects became busier~\cite{einstein1997aging}. 

Research has shown that prospective memory failures negatively affect day-to-day activities. For example, Haas et al.~\cite{haas2020} showed that more than 50\% of daily cognitive errors are linked to prospective memory impairments. In their study, general intentions suffered the most from prospective memory failures, including accomplishing domestic actions such as grocery shopping and watering plants, as well as work and study-related intentions, such as forgetting to print important documents or filling out forms. The same study showed that interpersonal intentions that involved other people were also impacted by prospective memory failures, and that older adults, who may prioritize social and emotional goals over instrumental and material goals, reported fewer errors compared to younger adults. 

Prospective memory can be augmented through time-based reminders (time-based prospective memory)~\cite{Barner2019TimeofdayEO} or cues in the environment (event-based prospective memory)~\cite{Einstein1990NormalAA}. Research has shown that the retention of prospective memory tasks is impacted by the cognitive load required to maintain the task, as well as age~\cite{Ballhausen2017}. More recent studies have explored the use of ``intention offloading,'' where external props and tools, such as calendars and diaries and strategically placed objects and technologies, such as smartphone alerts, are used to augment prospective memory~\cite{gilbert2015strategic,meyerhoff2021individual}. A recent review suggests that such intention offloading is highly effective and guided by metacognitive processes~\cite{gilbert2015strategic}. The authors conclude that metacognitive interventions, such as advice, encouragement, and feedback, could play an important role in helping people adapt to using cognitive tools to support memory~\cite{Gilbert2022OutsourcingMT}.

\subsection{Task Management Tools}

In efforts to help workers remember their tasks at appropriate times, including tasks initially encoded in emails, several researchers have designed a variety of task management tools~\cite{mukherjee2020smart,corston2004task,kokkalis2013emailvalet,bellotti_taskmaster,faulring2010agent,kramer2010pim}. Over the years, some researchers have proposed to entirely redesign the email interface~\cite{bellotti_taskmaster,faulring2010agent} while others have proposed modular add-ons for the email client~\cite{kramer2010pim}. Some researchers have leveraged AI methods to identify tasks automatically ~\cite{corston2004task,mukherjee2020smart} while others leverage crowdworkers~\cite{kokkalis2013emailvalet}. We provide more details on these email-based management tools and approaches below. 

In 1999, Horvitz, Jacobs, and Hovel described the Priorities system, which employed machine learning to learn to prioritize and to ideally alert users of unread email messages by considering multiple features drawn from the messages and an organization database, including language referring to project coordination, requests for assistance, and organizational relationships~\cite{Priorities1999}. Prioritized messages and notifications were presented in a distinct interface outside the users' main email system.  \citet{bellotti_taskmaster} presented a visionary system twenty-one years ago called \textit{TaskMaster} with the goal of redesigning the email interface to be optimal for managing tasks. Several years later, Faulring et al. proposed RADAR, which also proposed redesigning the email interface to be task-focused and mitigate email overload~\cite{faulring2010agent}. Along with modifying the view of the email inbox, RADAR leveraged machine learning to recognize emails that contain tasks and learn in what order tasks should be prioritized and completed. Another approach to helping workers remember their tasks effectively consists of understanding when and how to remind users about their tasks effectively. 

One task-focused email interface called SmartMail was designed to identify tasks within emails and summarize the task such that it can be quickly added to any to-do list~\cite{corston2004task}. Similarly, an application called Smart To-Do leveraged neural text generation and natural language processing to automatically generate to-do list items from email conversations~\cite{mukherjee2020smart}. Instead of using artificial intelligence to identify tasks within emails, an email client called EmailValet takes advantage of the crowd. EmailValet was designed to help minimize email overload and identify tasks while preserving user privacy~\cite{kokkalis2013emailvalet}. Instead of redesigning the entire email interface, Krämer proposes the To-Do Drawer, an add-on for Apple Mail ~\cite{kramer2010pim}. The To-Do Drawer allows users to drag and drop emails into the ``drawer'' to create tasks associated with the email conversation. The application is referred to as a ``drawer'' because it can be shown or hidden at any time in the Mail interface. 

\begin{figure*}[!h]
\includegraphics[trim=0 6cm 6cm 0cm,width=\linewidth,clip]{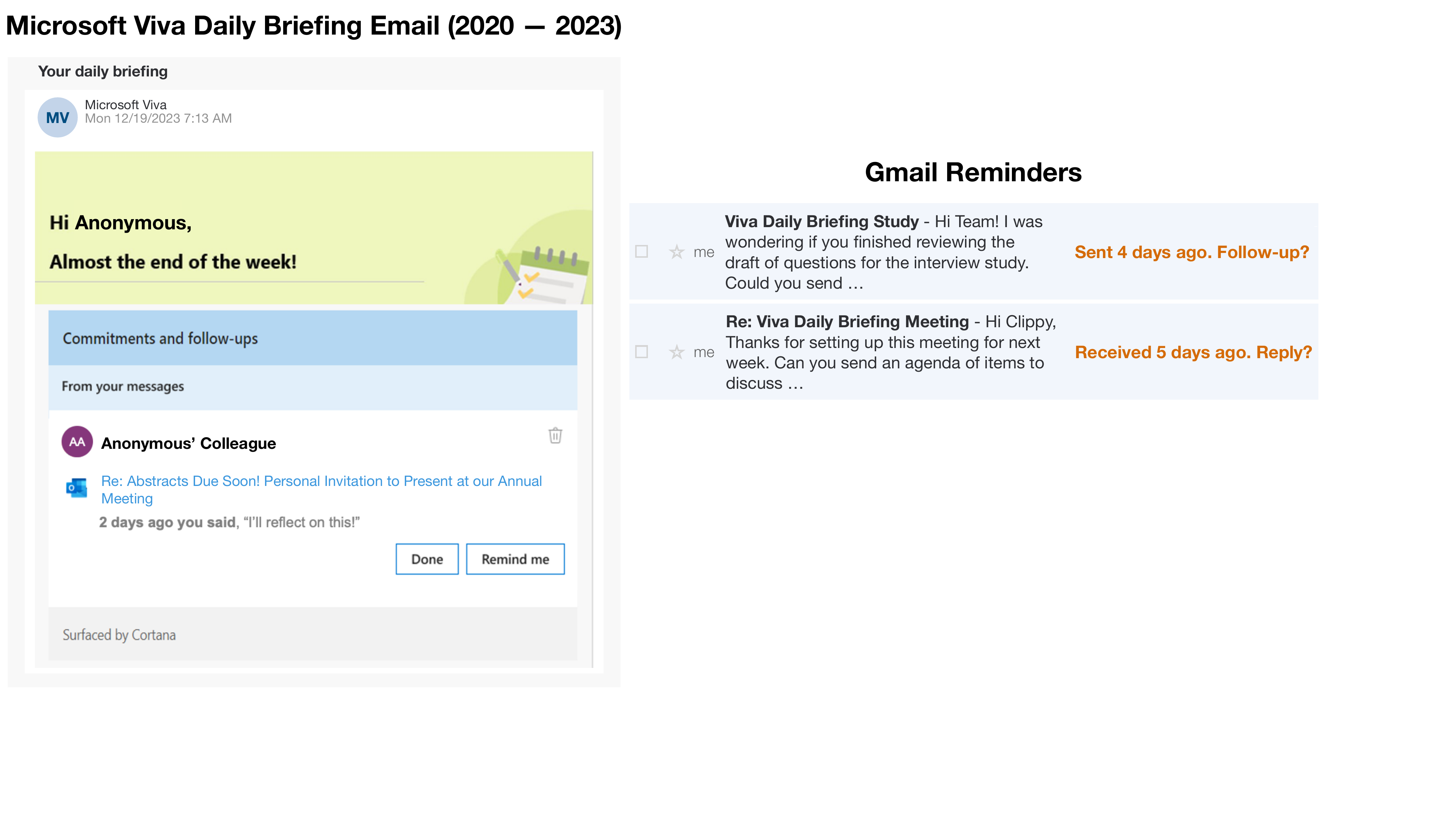}
    \caption{Example of a Viva Daily Briefing Email (2020 --- 2023) (left) and Gmail Reminders (right)}
    \label{fig:examples}
\end{figure*}

While all of the task management tools in the past have proposed redesigning the email interface or including an add-on in the email client, Microsoft and Google created intelligent task management tools that are integrated within the user's pre-existing workflow. Google's Gmail Reminders, shown in Figure \ref{fig:examples}, reminds users about emails they have not responded to yet (also known as requests) and emails that they had sent but have not yet been answered or acknowledged (also known as follow-ups) by showing orange text on the line of the suggested email at the top of their inbox~\cite{GoogleNudge}. 

Microsoft's Viva Daily Briefing Email (2020 --- 2023) was an email that was delivered to the user's email inbox every morning with emails associated with commitments, requests, and follow-ups. While the email contained additional sections at the time, we only focused on the ``Commitments and follow-ups'' section. \emph{Commitments} refer to something a user said they would do for someone, \emph{requests} as something that someone else is asking the user to do, and \emph{follow-ups} as something the user asked someone else to do. As seen in Figure \ref{fig:examples}, the Viva Briefing displays messages that the system has inferred contain commitments, requests, and follow-ups that the user has received and/or sent in Microsoft Outlook and Teams. The tool is directly connected to the user's Microsoft To-Do list---when a user clicks on the ``remind me'' button, the user can select a date and time to be reminded about the item. It will be added to their Microsoft To-Do list to be viewed outside of the Outlook client. When a user clicks ``done'', it will be marked as completed in Microsoft To-Do. 

\subsection{Human-AI Interaction with Task Management Tools}

Several studies have explored how people interact with intelligent assistants for task management in the workplace. One study analyzed how chatbots can be used for task management in the workplace and how participants interacted with it~\cite{10.1145/3173574.3173632}. Another study explores how information workers interact with a chatbot designed to improve worker productivity and well-being~\cite{kimani2019conversational}. Another study uses a grounded theory approach to present a framework on how intelligent assistants are generally used in the workplace~\cite{manseau2020ai} For example, the study reveals through interviews that workers used intelligent assistants for a variety of tasks including information acquisition and task automation~\cite{manseau2020ai}. In contrast, Khaokaew et al. seek to understand what people would desire from an intelligent assistant in the workplace~\cite{khaokaew2022imagining}. The authors focus on identifying desires related to the numerous capabilities of intelligent assistants in general. In contrast, we focus only on the task management functionality of the Viva Briefing. To date, the authors are not aware of any work studying how information workers interact with a widely fielded intelligent email-based task reminder system.

A recent study explores common categories of tasks in the workplace to inform the design of intelligent assistants for task management~\cite{trippas2019learning}. An earlier study conducted a survey and diary study to understand what types of tasks people want to remember and what tools they use to do so \cite{brewer2017remember}. One distinction between these two studies and our contributions is the focus of the surveys. \citet{trippas2019learning} focuses on thematically categorizing all work tasks that survey respondents provided, whereas we focus on understanding the work tasks that people tend to forget about. Unlike our study, \citet{brewer2017remember} does not focus strictly on work-related tasks. While both of these studies inform the design of intelligent assistants for task management, they do not study how users interact with current intelligent assistants that support asynchronous collaborations.

%% file: 03-experiences.tex
\section{Methods}

We capture users' experiences with the Viva Daily Briefing Email by analyzing the transcripts from 11 semi-structured interviews, open-ended responses from 11 pre-interview surveys, and survey responses from 45 additional participants to validate the interview findings. In this section, we describe the study design for the semi-structured interview and validating interview survey, along with how we analyzed the data.

\subsection{Semi-Structured Interviews}\label{ssi-recruit}

We recruited knowledge workers from Microsoft using a screening questionnaire that we designed. The screening questionnaire asked interested participants what their role was in the company, if they had an administrative assistant, and how often they used the Viva Daily Briefing Email. We grouped interested participants into four categories (\textit{i.e.}, frequent users, infrequent users, lapsed users, and rare/never users) based on their responses to how often they used the Viva Daily Briefing Email. We define \textit{frequent users} as those who use the system daily or a few times a week; \textit{infrequent users} as those who use it a few times a month; and a \textit{lapsed user} as someone who used to use the system frequently or infrequently but no longer does. To hear diverse perspectives, we selected a wide variety of knowledge workers with different roles within Microsoft and different usage patterns to interview. We tried to select participants such that our participant pool represented each usage type equally. We recruited 11 participants; however, we only had one lapsed user fill out the screening questionnaire, and they were not available to participate in the interview. We also only had two rare/never users fill out the questionnaire. 

All selected participants were required to complete a short, IRB-approved, pre-interview survey prior to their interview session. The pre-interview survey had at most $20$ questions; depending on the participant's responses, it could have had closer to 15 questions. The survey was a mix of open-ended response questions and Likert scale questions. The purpose of this pre-interview survey was to allow the interviewer to better understand the participant's work style and task management patterns throughout the workday, exposing how the Viva Daily Briefing Email complemented their existing workflow\footnote{See the screening questionnaire file and pre-interview survey questions file in the supplemental material.}. After completing the pre-interview survey, participants were directed to a booking system to schedule their interview session.

During the interview session, participants were asked to walk through some of their previous Viva Daily Briefing Emails and the tasks that were recommended to them in their emails. Participants were asked to describe the context and importance of the recommended task and recall if and when they took care of that task. Lastly, questions related to interacting with the recommended tasks were discussed. 
Participants who completed the interview were paid a $25$ USD Amazon Gift Card.

\subsection{Validating Interviews Survey}

The validating interviews survey is an IRB-approved survey that was designed to validate the findings from the semi-structured interviews across a broader population of knowledge workers within Microsoft \kmdelete{(part one)}. The survey was created and conducted after the semi-structured interviews were completed. This survey was designed to validate findings from the interviews, such as how people use the Viva Briefing and how much work style impacts how the user interacts with and benefits from the Viva Briefing \footnote{See the validating interviews survey file in the supplemental materials.}. Open-ended response questions and Likert scale questions were included in the survey. We sent the survey out to a broader population of knowledge workers within Microsoft who held roles similar to the participants from the semi-structured interviews. We received a total of 45 responses for the validating interviews survey. We sent our survey to employees that we randomly selected based on their role in the company. Participants took an average of $16.21$ minutes to complete the survey.

\subsection{Qualitative and Quantitative Analyses}

\paragraph{\textbf{Thematic Analysis}} We conduct a thematic analysis to draw insights from the open-ended responses in the pre-interview survey and the validating interview survey. A member of the research team identified initial themes and definitions of those themes. These themes and definitions were iterated until the two members of the research team responsible for coding the responses came to an agreement. After agreeing on the themes and definitions, the two coders coded all the responses. Each theme's definition is represented by a code that was created by the research team to code responses easily. The resulting themes and definitions are presented in Tables \ref{table: value} and \ref{table:things forgotten}.

\paragraph{\textbf{Regression Analysis}}
In order to do quantitative analyses, we defined two response metrics based on the survey questions. Some questions in the validating interviews survey consisted of several statements that can be paired together, which was the motivation for designing the metrics. We provide definitions for each metric below.

\textit{Memory of Low-Medium-Importance Tasks Metric}. This metric is the normalized sum of the coded 5-point Likert scale response to the following statements: ``I do not forget about low-importance tasks'' and ``I do not forget about medium-importance tasks'' where $1$ = \textit{strongly disagree}, $2$ = \textit{somewhat agree}, $3$ = \textit{neither agree nor disagree}, $4$ = \textit{somewhat agree}, and $5$ = \textit{strongly agree}. This metric represents a feature in the \texttt{Valuable} and \texttt{Interacts} models in Table \ref{table: value-interaction}.

\textit{Perceived Value Metric}. This metric is a normalized sum of the coded responses to two Likert scale questions from the survey. The first question, ``How valuable would you say that the Viva Daily Briefing Email is to you?'', presented the following options on a four-point Likert scale: $1$ = \textit{extremely not valuable}, $2$ = \textit{somewhat not valuable}, $4$ = \textit{somewhat valuable}, and $5$ = \textit{extremely valuable}. The value $3$ was not used as a code in order to represent a stronger distinction between somewhat not valuable and somewhat valuable. The second question asked the participant to select how they would be impacted if the ``Commitments and follow-up'' section was no longer available to them on a four-point Likert scale with the following options: $4$ = \textit{I would miss [it]}, $2.5$ = \textit{Neutral feelings about [it]}, $1$ = \textit{I would not miss [it]}, and $0$ = \textit{I don't know [this feature]}. This question was asked alongside a few other features that are included in the daily briefing email, which is why \textit{I don't know [this feature]} is provided as an option. This metric is used as the response variable for the \texttt{Valuable} model in Table \ref{table: value-interaction}.

\textit{Interaction Metric}. This metric was initially intended to be used as a feature in the \texttt{Valuable} model in Table \ref{table: value-interaction}. This feature proved to be the single most important feature in the \texttt{Valuable} model. Therefore, we chose to use this metric as the response variable in a second model, \texttt{Interacts}, to see how the other features are associated with positively interacting with the reminder feature. Observations from our semi-structured interviews led us to define six different ways that encompass how users interact with this task reminder system: (1) Users briefly skim the surfaced information, (2) click the buttons to interact with the surfaced information, (3) closely read everything surfaced, and/or (4) closely read only certain things surfaced. We also learned that users may (5) open the email and not read it just to clear the email notification, while others (6) do not open it at all. These last two interactions are denoted as negative interactions, while the first four are denoted as positive interactions. Participants were shown all six interactions and asked to select all the ways that they interacted with the email. For each interaction selected, participants were asked how often they interact with the email in that way on a five-point Likert scale with the following options: $5$ = \textit{daily}, $4$ = \textit{a few times a week}, $3$ = \textit{a few times a month}, $2$ = \textit{I used to do this but no longer do}, $1$ = \textit{rarely to never}, and $0$ = if the participant did not select this interaction. We calculate the normalized sum of the four positive and two negative interactions with values ranging from $-0.5$ to $1.0$. \kmdelete{This metric is used as a feature in the \texttt{Valuable} model and as the response variable for the \texttt{Interacts} model in Table \ref{table: value-interaction}.}

\textit{Independent Variables}. A majority of the independent variables in both models are based on the coded five-point Likert scale responses from strongly disagree to strongly agree. We excluded certain features from the model that had significant, positive correlations with another feature being considered. The following features had a statistically significant ($p<0.01$), moderately high ($> 0.5$) correlation with another feature in the data set: \textit{organized}, \textit{using email as a task manager}, and \textit{ad-hoc meetings}. Thus, the features used in the models that follow this Likert scale include \textit{clean \& compact email inbox}, \textit{categorizes/files emails}, \textit{receive a lot of emails w/ projects/tasks}, \textit{creates/communicates about tasks via email}, \textit{involved in many distinct projects}, \textit{many scheduled meetings}, \textit{procrastinates}, and \textit{delegates tasks}. However, some of the other features being used in the models do not follow the same 5-point Likert scale.

\section{Results} 

We answer our four research questions through a mixed-methods approach. We take note of how interview participants incorporated the Viva Daily Briefing Emails into their workflow and conducted a thematic analysis of open-ended responses from the surveys about why the tool was valuable. These findings are elaborated in Section \ref{rq1-res} to answer \textbf{RQ1}. In Section \ref{rq2-res}, we answer \textbf{RQ2} with a second thematic analysis on open-ended responses from the pre-interview and validating interview surveys to understand the type of information people would like AI-powered reminders to remind them about. We take a quantitative approach to \textbf{RQ3} in Section \ref{rq3-res} by doing a regression analysis on self-reported workstyles and the perceived value of Viva Daily Briefing Email. Lastly, we provide insight into \textbf{RQ4} in Section \ref{rq4-res}. We summarize how knowledge workers would like to interact with AI-powered reminders by detailing how the interview participants wanted to interact with their previously recommended tasks.

\paragraph{Participant Statistics} Overall, we interviewed five frequent users, four infrequent users, and two rare/never users. We interviewed 11 Microsoft employees for an average of $49.01$ minutes (sd = 5.60 minutes), and the pre-interview survey took an average of $15.23$ minutes (sd = $10$ minutes). The following roles were represented across the 11 employees that we interviewed: one software developer/engineer, two program/product managers, two hardware engineers, two user experience researchers, two business admins, one manager, and one writer. None of the participants had executive assistants.

The participants who completed the validating interview survey were not required to provide their role at Microsoft. We also do not categorize these participants the same way we did for the interview participants, as we learned from the interviews that there is no clear metric to determine a frequent versus infrequent user.

\subsection{RQ1: Incorporating AI-Powered Reminders into Workflows} 
\label{rq1-res}


During the semi-structured interviews, we learned that participants incorporated this tool into their workflow in very different ways, which was based on their level of interaction with the recommended tasks, perception of the accuracy of the recommendations, and the amount of communication about asynchronous collaborations. Some of the knowledge workers passively incorporated the tool into their workflow because they did not communicate frequently about asynchronous collaborations via Outlook or Teams, while the tool played a more active role in the workflow for other participants who exchanged a lot of communication related to their asynchronous collaborations. For example, \textit{P4}, categorized as a frequent user, heavily relies on Outlook and Teams to asynchronously communicate about collaborations and has actively incorporated AI-powered reminders into their workflow. Compared to the other participants, \textit{P4} had numerous suggested reminders in their Viva Daily Briefing Email. \textit{P4} noted how they open the Viva Daily Briefing Email first thing every morning and will leave it visible in a window on their desktop throughout the day as they attend meetings and complete their work. This participant described how the AI-powered reminders influenced their workflow and asynchronous collaborations by causing them to adjust how they send emails to collaborators: ``\textit{... any action items I'll put in the body of the e-mail, so that way it would get caught [\textit{by the viva daily briefing email}]. So it would show here [\textit{in the next viva daily briefing email}]}.'' Another frequent user, \textit{P11}, supports the business and administration needs of a group larger than 50 people, resulting in a high volume of communication and collaboration over Teams and Outlook. \textit{P11}\kmdelete{, also categorized as a frequent user,} noted that the Daily Briefing Email ``\textit{... has changed a lot of how I've done my day to day}.'' \textit{P11}'s workflow now consists of using AI-powered reminders to help them generate their own To-Do list in a separate text document that they refer to throughout the day.

The Daily Briefing Email did not have such an influential impact on everyone's asynchronous collaborations. For example, \textit{P1}, who was deemed as a frequent user, noted that they ``\textit{... go through it and soak in the information and use the information as needed,}'' which is a passive way to incorporate AI-powered reminders into the workflow. This tool minimally impacted their asynchronous collaborations as \textit{P1} says they ``\textit{... don't make a lot of commitments...[and] don't have a whole lot of things to follow up on.}'' Similarly \textit{P2}, \textit{P6}, and \textit{P10} ``\textit{... will just scroll through the things.}'' \textit{P2} noted that they rarely asynchronously communicate with their team about collaborations: ``\textit{A lot of time it's we just call each other on teams...if the person is free we just call and sort out the issue.}'' \textit{P6}, categorized as an infrequent user, noted that they treat the reminders to be ``\textit{... almost like a fail-safe to making sure that I don't miss anything.}''

Aside from how the AI-powered reminders were incorporated into workflows and impacted asynchronous collaborations, we look at the value the tool brings through a thematic analysis to see why people incorporated the tool into their workflow. To investigate this, the interview participants were asked in the pre-interview survey if the Viva Daily Briefing Email was useful (yes or no). In the validating interview survey, we also asked participants a similar question on a four-point Likert scale: ``\textit{How valuable is the Viva Daily Briefing Email to you?}''. Both groups were asked to elaborate briefly on their answer. The final coding of the open-ended responses in Table \ref{table: value} had a high inter-rater reliability (IRR) score of $90.38\%$ for the validating interview survey responses and $77.27\%$ for the pre-interview survey responses. The survey count in Table \ref{table: value} represents the number of participants from the validating interview survey, and the interview count represents the number of participants from the pre-interview survey who had a response that matched one of the themes. We provide further details about each theme below and representative quotes from the responses.

\begin{table}[ht!]
\centering
\begin{tabular}{l p{0.38\linewidth} c c} 

 \hline
 Code & Definition & Survey Count (N=45) & Interview Count (N=11)\\ [0.5ex] 
 \hline
SUP & Staying up to date on tasks and meetings; summary of meetings and tasks & $7$ $(15.56\%)$ & $3$ $(27.27\%)$\\
  RFER  & Reminder about forgotten tasks; extra reminder for tasks, meetings, etc. & $16$ $(35.56\%)$ & $2$ $(18.18\%)$\\
 BCM & Book time on calendar & $4$ $(8.89\%)$& $1$ $(9.09\%)$ \\
 ALKN & Already aware of the information and/or taken care of & $9$ $(20.00\%)$& $2$ $(18.18\%)$ \\ 
 NOA & Not presenting accurate or most up-to-date information & $12$ $(26.67\%)$ & $2$ $(18.18\%)$ \\[1ex] 
 \hline
\end{tabular}
\caption{\textbf{The value of the Viva Daily Briefing Email in the workplace.} The open-ended responses from the validating interview survey question about value were coded into themes and are shown as the survey count. The responses from the pre-interview survey question about value were also coded into themes and are shown as the interview count.}
\Description{A table with four columns and seven rows. Each row represents a different theme that we identified from our thematic analysis. The first column is the code for the theme, the second column is the definition of the theme, the third column is the number of participants from the survey, and the fourth column is the number of participants from the interview. }
\label{table: value}
\end{table}

\textit{\textbf{RFER: An extra reminder or reminder of forgotten tasks and meetings.}} The highest number of participants commented on how the Viva Briefing provided them with an extra reminder or reminder of forgotten tasks or meetings. Of the $16$ participants from the broader survey, only one participant who made a statement about this also ranked the briefing email as \textit{somewhat not valuable} to them. $12$ participants ranked the briefing email as \textit{somewhat valuable} and three ranked it as \textit{extremely valuable} to them. Both participants from the pre-interview survey who made this statement viewed this tool as useful, which we used as a proxy for value. One of the participants who found the Viva Briefing \textit{extremely valuable} said, ``\textit{I find the briefing email extremely valuable because [it] help [sic] me remain on track with tasks and engagements that I might have inadvertently forgotten about and/or failed to register in my own `to do' list}.'' One participant who found the Viva Briefing \textit{somewhat valuable} describes how the reminders positively impact their asynchronous collaborations by reminding them to ``\textit{follow up on certain email/messages I may have sent ... or ... reply back to people who asked something of me and I had indicated to them that I would get back to them shortly}.'' Similarly, a participant who views the briefing email as \textit{somewhat valuable} said, ``... \textit{Reminds me of some tasks that I need to complete that I might have forgotten about}.'' 

\textit{\textbf{ALKN: Already aware of the information and/or already taken care of.}} While many participants made comments about how this system is \textit{somewhat} or \textit{extremely valuable} to them, we also received a lot of comments on why this system is \textit{somewhat not} or \textit{extremely not valuable}. Of the nine survey participants who commented on this theme, five said that this tool is \textit{extremely not valuable}, and three said that it is \textit{somewhat not valuable}. One participant said that it is \textit{somewhat valuable}. Both interview participants who commented on this viewed the tool as not useful. One participant from the survey who views this tool as \textit{somewhat not valuable} said, ``\textit{It tries to [show] outstanding tasks of this [sic] that I need to respond to but I have already completed or tracked separately elsewhere}.'' Similarly, a participant who sees this tool as \textit{extremely not valuable} said, ``\textit{...all I see if [sic] reminders for things already done}.''

\textit{\textbf{NOA: Not presenting accurate or the most up-to-date information.}} This was the second most frequent statement we saw behind \textbf{\textit{RFER}}. Out of the $12$ participants from the broader survey that commented on the tool's inaccuracy, three participants said the system is \textit{extremely not valuable}, six said it is \textit{somewhat not valuable}, and three said it is \textit{somewhat valuable} to them. Both interview participants who commented on this said that the briefing email was not useful. A participant who said that the briefing email is \textit{somewhat not valuable} said, ``\textit{Often the recommendations it provides of suggested tasks are inaccurate and not helpful}.'' A participant who said that the tool is \textit{extremely not valuable} commented, ``\textit{Never find info on it that aligns with what I'm working on}.''

Overall, we observed that knowledge workers incorporated AI-powered reminders into their workflow in passive and active ways to support their asynchronous collaborations. We also learned that knowledge workers found value in being reminded of forgotten tasks and receiving extra reminders (\textbf{\textit{RFER}}). They did not find value in AI-powered reminders presenting inaccurate or out-of-date information (\textbf{\textit{NOA}}) and presenting information that they already know (\textbf{\textit{ALKN}}). Since the largest number of participants made \textbf{\textit{RFER}} comments and simultaneously saw the Viva Briefing as valuable, we conducted a second thematic analysis to uncover what tasks are forgotten about. 

\subsection{RQ2: Characterizing Information that AI-Powered Reminders Should Show} 
\label{rq2-res}

With the highest number of participants valuing the briefing email for providing an extra reminder or reminders of forgotten tasks, we conduct a second thematic analysis to characterize what types of tasks are forgotten about. We asked participants in the broader survey to, ``Describe the kinds of work-related commitments that you are most likely to forget about''. Below, we provide detailed descriptions of each theme with at least eight participants and some representative quotes from the responses. 

\begin{table}[h!]
\centering
\begin{tabular}{l p{0.6\linewidth} c} 
 \hline
 Code & Definition & Survey Count (N=45) \\ [0.5ex] 
 \hline
DNF & Does not forget about things  & $4$ $(8.89\%)$ \\
  LIT  & Forgets about tasks or meetings that are not of high importance or urgency & $11$ $(24.44\%)$ \\
  DWD & Forgot about task or meeting because it was not written down & $2$ $(4.44\%)$\\
 ADHC & Forgets about ad hoc meetings/tasks; meetings that you do not need to participate in & $10$ $(22.22\%)$   \\
 NWT & Forgets about non-project-related tasks or meetings & $5$ $(11.11\%)$ \\
 SML & Forgets about tasks that require little effort & $8$ $(17.78\%)$ \\ 
 VER & Forgets things that were verbally said & $2$ $(4.44\%)$  \\
 NA & Not applicable; could not code response based on identified themes or could not agree on code for response & $7$ $(15.56\%)$ \\[1ex] 
 \hline
\end{tabular}
\caption{\textbf{Information that tends to be forgotten.} Thematic Analysis of the responses to the question ``Describe the types of work-related commitments that you are most likely to forget about'' from the survey.}
\Description{A table with three columns and seven rows. Each row represents a different theme that we identified from our thematic analysis. The first column is the code for the theme, the second column is the definition of the theme, and the third column is the number of participants from the survey.}
\label{table:things forgotten}
\end{table}

\textit{\textbf{LIT: Forgets about tasks and meetings that are not urgent or of high importance.}} Of the $11$ participants who commented on forgetting tasks that are not urgent or of high importance, six participants found this tool \textit{somewhat valuable}, and one participant found it \textit{extremely valuable}. On the contrary, three found the tool to be \textit{extremely not valuable}, and one found it to be \textit{somewhat not valuable}. The four participants who found this tool to be \textit{somewhat not} or \textit{extremely not valuable} also stated that the tool was not accurate (\textit{\textbf{NOA}}) or simply useless (\textbf{\textit{NOV}}). Of the participants who found this tool to be \textit{somewhat} or \textit{extremely valuable}, commented on forgetting tasks such as, ``\textit{Replying to non-urgent emails that take time}'' and ``\textit{Following up on a low priority request which may not be due till much later}.''

\textit{\textbf{ADHC: Forgets about ad hoc meetings or tasks, or non-participatory meetings.}} $10$ participants mentioned that they forget about ad hoc tasks or meetings, especially meetings that do not require participation. Of the $10$ participants, five rated the tool as \textit{somewhat not valuable} to them, while four rated the tool as \textit{somewhat valuable} and one rated \textit{extremely valuable}. A participant who views this tool as \textit{somewhat not valuable} said that they tend to forget about ``\textit{...ad hoc requests via email that do not conform to my already established structure of tasks}.'' Many of the participants' responses simply stated, ``\textit{ad hoc meetings}'' or ``\textit{ad hoc requests}''.

\textit{\textbf{SML: Forgets about tasks that require little effort.}} The third most popular type of information that our participants said they tend to forget about is small tasks. Of the $8$ participants who said this, three view this tool as \textit{somewhat valuable} and three view it as \textit{somewhat not valuable}. One participant sees this tool as \textit{extremely not valuable} while the other sees it as \textit{extremely valuable}. These participants describe the tasks that require little effort to include: ``\textit{Sending a message to someone, setting up a meeting with someone, sharing a link to a paper if someone asked for one}'', ``\textit{Little tasks that get lost in the emails}'', and ``\textit{Follow up on an email, reply to a thread}''. These tasks, despite requiring little effort, can still impact asynchronous collaborations if they are not addressed in a timely manner.

This thematic analysis aimed to identify what types of work-related tasks people forget about. When participants were asked about the types of work-related tasks they forget about, the top three themes that emerged were \textbf{\textit{LIT}} (forgetting about tasks that are not of high importance or urgency), \textbf{\textit{ADHC}} (forgetting about ad hoc tasks or meetings that do not require participation), and \textbf{\textit{SML}} (forgetting about tasks that require little effort). Through quotes from the participants, we provide examples of what each of these tasks could be for someone. 

%% file: 04-work-styles.tex



\subsection{RQ3: Impact of Workstyle on the Value of AI-Powered Reminders}
\label{rq3-res}

\begin{table}[b!] \centering 
\begin{tabular}{l c c } 
\hline 
\textit{Independent variables:} & \multicolumn{2}{c}{\textit{Dependent variable:}} \\ 
\hline
\\[-1.8ex] & \multicolumn{1}{c}{Valuable} & \multicolumn{1}{c}{Interacts} \\ 
\hline \\[-1.8ex] 
(Intercept) & $0.730^{*}$ & $0.078$ \\
 Clean \& compact email inbox & $-0.022$ & $-0.012$ \\ 
  Categorizes/files emails & $0.009$ & $0.041$ \\ 
  Receive a lot of emails w/ projects/tasks & $-0.006$ & $-0.035$ \\ 
  Communicates about tasks via email & $0.010$ & $0.127^{*}$ \\ 
  Creates tasks from emails & $0.011$ & $0.077^{*}$ \\ 
  Involved in many distinct projects & $0.020$ & $0.007$ \\ 
  Many scheduled meetings & $-0.016$ & $-0.108^{*}$ \\ 
  Procrastinates & $-0.045$ & $0.023$ \\ 
  Delegates tasks & $0.021$ & $-0.084^{*}$ \\ 
  Low\_Medium\_Norm & $-0.186$ & $-0.114$ \\ 
  Interactions & $0.444^{**}$ & \textit{---} \\  
 \hline \\[-1.8ex] 
N & \multicolumn{1}{c}{45} & \multicolumn{1}{c}{45} \\ 
R$^{2}$ & \multicolumn{1}{c}{$0.484$} & \multicolumn{1}{c}{$0.459$} \\
Adjusted R$^{2}$ & \multicolumn{1}{c}{$0.312$} & \multicolumn{1}{c}{$0.300$} \\ 
Residual Std. Error & \multicolumn{1}{c}{$0.224$ (df = $33$)} & \multicolumn{1}{c}{$0.247$ (df = $34$)} \\ 
F Statistic & \multicolumn{1}{c}{$2.812$ (df = $11$; $33$)} & \multicolumn{1}{c}{$2.885$ (df = $10$; $34$)} \\ 
\hline 
\hline \\[-1.8ex] 
\textit{Note:}  & \multicolumn{2}{r}{$^{*}p<0.05$; $^{**}p<0.01$} \\ 
\end{tabular} 
\caption{People who reported that the Viva Briefing is valuable to them are people who positively engaged with it, such as clicking the buttons and closely reading the surfaced information (\texttt{Valuable} model). Therefore, we identify features that are associated with positively interacting with the Viva Briefing (\texttt{Interacts} model). People who self-reported to positively interact with the Viva Briefing are those who communicate about their tasks via email, create tasks from emails, do not have many meetings, and do not tend to delegate their tasks.} 
\label{table: value-interaction} 
\Description{A table with three columns where the first column represents the independent variables and the second two columns represent dependent variables. Two linear models were created, which is why there are two dependent variable columns. There is a list of all the features used in the rows and their coefficients for each model. Summary statistics of the model are presented in the last five rows. }
\end{table} 

We perform quantitative analyses of responses to self-assessment questions in the validating interviews survey\footnote{Survey questions provided in the supplemental material.} to identify which work styles benefit from this system and how their work style affects the perceived value of the briefing email (\textbf{RQ3}). We regressed on features that are based on the survey questions to create two models. 

To identify who values the briefing email, we created two linear regression models shown in Table \ref{table: value-interaction}. These models allow us to see if there is an association between the self-reported value of the tool and different work styles. We use someone's work and task management patterns to define their work style. We defined these patterns based on patterns that we observed during the semi-structured interviews and findings from previous works~\cite{dabbish2006email,whittaker1996email}.

The only statistically significant feature that is associated with someone finding the Viva Briefing valuable (model \texttt{Valuable} in Table \ref{table: value-interaction}) is the \texttt{Interaction} feature. As users positively interact with the service, they find it more valuable, which is expected. However, this provides little detail about the work style of people who positively interact with this tool. 

We created a second model (\texttt{Interacts}) to see if we can identify any features that are associated with someone who would interact with this tool to infer which work styles view the tool as valuable. 
By doing so, we see four specific features associated with workstyle to be statistically significant. We find that people who communicate about their tasks via email ($p < 0.05$), create tasks from emails that they receive or send ($p < 0.05$), do not have many scheduled meetings ($p < 0.05$), and do not delegate their tasks to others ($p < 0.05$) are people who self-report to positively interact with this tool. This trend can be corroborated with our qualitative findings presented in Section \ref{rq1-res}.

%% file: 05-interactions.tex
\subsection{RQ4: The Future of Interacting with AI-Powered Reminders}
\label{rq4-res}

We observed that the participants who found value in the Viva Daily Briefing Email were those who positively interacted with it (\textit{i.e.}, reading the email and interacting with the recommended tasks). However, the possible interactions with the recommended tasks are limited. The technology behind the Viva Daily Briefing Email has a lot of potential. Perhaps if different interactions were offered, we would see different trends. In order to leverage this technology to its fullest potential, we need to better understand how people would want to interact with the information that is surfaced for them. We open up the discussion of what it could look like for workers to interact with AI-powered reminders more broadly by soliciting how knowledge workers would like to interact with the tasks recommended by the Viva Daily Briefing Email. More specifically, what do people want to be able to do with AI-powered reminders, and how do they want to be able to do it? We describe the common themes that emerged from the retrospective analysis portion of the semi-structured interviews below.

Currently, the AI-powered reminders in the Viva Daily Briefing Email offer four primary interactions, as shown in Figure \ref{fig:examples}. The user can click the ``Done'' button, the ``Remind me'' button, the trashcan icon, or the email subject line. We learned from the retrospective discussion during the semi-structured interviews that not all participants know what these buttons do. Furthermore, we learned that these buttons do not reflect the types of ways people would like to interact with the AI-powered reminders of recommended tasks. 

Similarly to \citet{bellotti_taskmaster}, we observed that tasks fit broadly into one of three categories: \textit{rapid-response, extended-response, and interdependent} tasks. For some tasks, participants wanted to be able to respond quickly within the Viva Daily Briefing Email (\textit{rapid-response tasks}), while other tasks required \textit{extended} effort to complete. \textit{P10} really wanted an in-line, quick reply functionality where they could reply to the recommended tasks in emails because they ``\textit{... wanna stay in that e-mail environment...}'', they ``\textit{do not wanna switch contexts.}'' \textit{P1}, \textit{P2}, and \textit{P11} shared similar sentiments.

For some \textit{extended-response} tasks, participants wanted to schedule a time in their calendar to take care of it, which we refer to as \textit{scheduled-response} tasks. This desire was also observed by~\citet{khaokaew2022imagining}. \textit{P10} expressed this desire in more detail: ``\textit{... the more useful thing would be if I could schedule it in my calendar because sometimes it [Viva Daily Briefing Email] will tell me something, and I just wanna click and go, okay, I don't have time to do that right now. But, go ahead and find time on my calendar...}.'' \textit{P10} suggests that when AI-powered reminders recommend a task that will take a lot of time and cannot be forgotten about, there should be an easier way to schedule these in the moment of review. This functionality is different than the current ``Remind me'' button, which only reminds the user about the task at a set time.

Alternatively, some participants did not want to decide in the moment when they would take care of a task~\cite{kramer2010pim}, but they wanted to quickly and easily save it somewhere to refer to later so that they could check back on its progress and not forget about it. \textit{P2} remarked, ``\textit{If I can create small tasks and move tasks around in a tab in Outlook would be very helpful.}'' For these types of tasks, you do not want to decide at the moment of review when to be ``reminded''.  We refer to these types of tasks as \textit{undetermined-response} tasks. Lastly, participants wanted a way to acknowledge tasks even though there was no current action required from them, which we call \textit{no-response-needed} tasks.

%% file: 06-discussion.tex
\section{Discussion}\label{discussion}

This work explores how a widely deployed AI-powered reminder system, Microsoft's Viva Daily Briefing Email, was utilized by knowledge workers at Microsoft. Through a variety of methods, including semi-structured interviews and surveys, we apply mixed methods to analyze the experiences and futures of AI-powered reminders for collaborative tasks. We discuss the implications of our findings.

\textbf{AI-powered reminders were incorporated into workflows differently, and knowledge workers wanted to be reminded about forgotten tasks.}
People establish their own means of managing tasks and triaging emails from asynchronous collaborations. Many people keep some sort of ongoing to-do list filled with tasks that they want to remember to complete later. Some of these tasks are organized based on the project, while some are organized based on urgency. While some users felt the Viva Daily Briefing Email helped them manage their asynchronous collaborations, not all participants found value in using it throughout the workday. Some participants passively acknowledged the recommended tasks, while others adapted their workflow, such as when and how they managed tasks from the Viva Briefing. We observed that the extent of participants' asynchronous communication regarding their collaborations could affect how impactful AI-powered reminders are to their collaborations. Some participants also reported that the Viva Briefing surfaced tasks that they already had on their own to-do list or that they had already completed. The briefing email was designed to prioritize surfacing tasks that have a higher likelihood of being urgent. However, we've discovered that people do not necessarily want to be reminded about something they would have on their own to-do list or is of high importance or urgency. People found value in the Viva Daily Briefing Email reminding them about tasks that they forgot about or missed. People self-reported forgetting about tasks that are not of high importance or urgency but still need to be taken care of eventually. Tasks of higher importance have been shown to be added to some form of a to-do list (excluding repetitive tasks) \cite{10.1145/985692.985785}. Similarly, Walter and Meier's review shows that the importance of a task positively contributes to one's ability to remember to do something in the future \cite{10.3389/fpsyg.2014.00657}. These two studies provide insight as to why people may be more prone to forgetting about low- and medium-importance tasks. Furthermore, Horvitz et al. observed that atypical meeting locations and attendees and the presence of managers or senior leaders in an organization were linked to higher event memorability, while recurrent events in typical locations were linked to events assessed as lower memorability \cite{horvitz2004learning}. Future designs of AI-powered reminders should reconsider whether to prioritize surfacing tasks of high importance and urgency or tasks that are more likely to be forgotten about. Conducting longitudinal studies would provide better insight into how active versus passive incorporation of AI-powered reminders in workflows impacts the outcome of asynchronous collaborations. Understanding how knowledge workers interact with AI-powered reminders can provide businesses and leaders with a new toolkit to help augment their teams by helping members avoid forgetting about deadlines and tasks.

\textbf{Workstyle impacts the perceived value of AI-powered reminders in the workplace.} 
Our quantitative analyses found that people who positively interact with the daily briefing email are most likely to find value in it. This suggests a correlation between people who interact with the daily briefing email and people who value it. While this correlation is to be expected, our interviews revealed that people who positively interact with the system are not the only ones who find value in the system. We regressed several work and task management features on the \textit{Interaction Metric} to infer other predictors for finding the briefing email valuable. People who communicate about their tasks via email, create tasks from emails, do not have many scheduled meetings, and do not delegate tasks are associated with positively interacting with the daily briefing email. One previous study discusses how being able to delegate tasks is a sign of being in a higher role or ``feeling in control'' \cite{mackay1988more}. Employees in managerial positions have more control over the requests to accept or deny and may likely have a secretary to manage many of these requests \cite{mackay1988more}. Considering our results, we suggest that those who are not in a managerial position and/or do not have an administrative assistant may find more value in the current email-based task reminder system. While insignificant, one participant from our semi-structured interview study was a manager and a rare/never user of this task reminder system. AI-powered reminders may want to consider the role of the end-user in their organization, how many collaborations they have, and whether they are primarily asynchronous or synchronous collaborations. For managers, AI-powered reminders may be more valuable if they are able to distinguish between tasks that will most likely be taken care of by an administrative assistant and tasks that cannot be taken care of by an administrative assistant. This classification could help AI-powered reminders be more meaningful to users who have administrative assistants and, as a result, may be incorporated into workflows differently.

\textbf{Knowledge workers want to interact with AI-powered reminders in ways that augment their workflow.} Determining what interactions to incorporate into AI-powered reminder systems requires developing an understanding of the users of the tool---and the models powering that tool. Through our semi-structured interviews, we identified different ways that participants wanted to interact with the AI-powered reminders but currently could not due to the design constraints of the briefing email. Knowledge workers wanted to interact with recommended tasks in ways that would augment their workflow, such as by directly scheduling time in their calendar to address the recommended task or quickly replying to the recommended task within context. We also learned that the current interactions available in the tool are ambiguous to the users, which discourages some users from interacting with the tool and, as a result, viewing it as having no value to them. Our quantitative findings show that not all work styles currently benefit from or find value in the Viva Briefing. However, it is not necessarily the case that those other work styles would never benefit from or value AI-powered reminder systems if they were more personalized to their individual goals, workflows, and values in the workplace. These goals and values may change over time as people change projects, roles, or teams, motivating the need for AI-powered reminder systems to be able to learn the user's current work style and adapt to it. Many asynchronous collaborations, as we confirmed from our interview participants, can happen in other channels of communication, such as on GitHub, video calls, in-person interactions, or other project management tools. This observation is corroborated by qualitative findings from~\cite{khaokaew2022imagining}. With asynchronous collaborations happening on various communication channels, the design of AI-powered reminders will have to consider how to parse numerous communication channels to improve the value of the surfaced information. Designing AI-powered reminders with interactions that are more aligned with how knowledge workers want to augment their workflow and asynchronous collaborations would provide a lot of value to businesses and teams, especially as we continue to embrace co-located and remote work environments. 




%% file: 07-limitations.tex
\section{Limitations}
We briefly elaborate on points of limitations within our study design and analyses that may impact the generalizability of our findings. 

\textbf{Single AI-powered reminder system.} This work focuses on evaluating the once widely deployed AI-powered reminder system from Microsoft, called the Viva Daily Briefing Email. This AI-powered reminder system was designed within an email. However, AI-powered reminder systems can take on many different forms aside from email. Given that AI-powered reminders can exist in different formats, it is unclear to what extent our findings generalize to non-email-based AI-powered reminders. This limitation should be taken into consideration when interpreting the implications of our findings. 

\textbf{Self-assessment methodology.} Our surveys and interviews relied heavily on self-assessments which add subjectivity to our findings, especially the statistical analyses we conducted to answer how workstyle impacts the perceived value of AI-powered reminders. The self-assessment nevertheless surfaced findings that shed light on the use of the system and that frame new questions and research directions that can guide the refinement of AI-powered reminders in the future.

\textbf{Few managers and workers with administrative assistants.} We reached out to a broad swath of employees but were limited in our ability to engage with managers and other workers who had administrative assistants. Only one manager ended up participating and they did not have an administrative assistant. This limitation should be taken into consideration when considering our findings. Future research should prioritize recruiting subjects from this population. 


%% file: 08-conclusion.tex
\section{Conclusion}


People use a variety of communication channels to maintain asynchronous collaborations. Language models and other AI methods are increasingly being used to aid people in identifying and prioritizing their tasks from asynchronous collaborations, yet we know little about how these tools are being used and by whom. We studied the usage and perceived value of a widely fielded email-based task reminder system, Microsoft's Viva Daily Briefing Email, as a vehicle to understand how AI-powered reminders have been incorporated into workflows. We identified how knowledge workers have incorporated AI-powered reminders into their workflow and examined the value the reminders bring to asynchronous collaboration in the workplace. We also pursued insights about the type of information that knowledge workers prefer to be reminded about via AI-powered reminders. We further explored the characteristics of users who express that they value and benefit from the tool. Finally, we engaged with participants about how they would like to interact with AI-powered reminders in future systems. Our qualitative results show that people found value in being reminded about tasks that they forgot about or missed. More generally, our interviews revealed that more work needs to be done to characterize the value of AI-powered reminders for collaborative tasks in the workplace. Our quantitative results showed that people who keep their email inbox clean and are involved in many distinct projects may find less benefit in the daily briefing email than others.  Our multipronged study and results provide new insights into how knowledge workers interact today with AI-powered reminders and their aspirations for engaging with future AI systems designed to support task management and collaboration. Moving forward, we see this work as a stepping stone to further research on AI-powered reminders in the workplace to support asynchronous collaborations. We see great opportunities ahead to bring together researchers in design, task management, and AI to develop and refine computing systems that can deliver on the dream of augmenting human memory, reasoning, and collaboration.

%% file: main.bbl

\begin{thebibliography}{42}


\ifx \showCODEN    \undefined \def \showCODEN     #1{\unskip}     \fi
\ifx \showDOI      \undefined \def \showDOI       #1{#1}\fi
\ifx \showISBNx    \undefined \def \showISBNx     #1{\unskip}     \fi
\ifx \showISBNxiii \undefined \def \showISBNxiii  #1{\unskip}     \fi
\ifx \showISSN     \undefined \def \showISSN      #1{\unskip}     \fi
\ifx \showLCCN     \undefined \def \showLCCN      #1{\unskip}     \fi
\ifx \shownote     \undefined \def \shownote      #1{#1}          \fi
\ifx \showarticletitle \undefined \def \showarticletitle #1{#1}   \fi
\ifx \showURL      \undefined \def \showURL       {\relax}        \fi
\providecommand\bibfield[2]{#2}
\providecommand\bibinfo[2]{#2}
\providecommand\natexlab[1]{#1}
\providecommand\showeprint[2][]{arXiv:#2}

\bibitem[Goo(2018)]%
        {GoogleNudge}
 \bibinfo{year}{2018}\natexlab{}.
\newblock \bibinfo{title}{Gmail will now remind you to respond}.
\newblock
\newblock
\urldef\tempurl%
\url{http://workspaceupdates.googleblog.com/2018/05/gmail-remind-respond.html}
\showURL{%
\tempurl}


\bibitem[Ballhausen et~al\mbox{.}(2017)]%
        {Ballhausen2017}
\bibfield{author}{\bibinfo{person}{Nicola Ballhausen}, \bibinfo{person}{Katharina Schnitzspahn}, \bibinfo{person}{Sebastian Horn}, {and} \bibinfo{person}{Matthias Kliegel}.} \bibinfo{year}{2017}\natexlab{}.
\newblock \showarticletitle{The interplay of intention maintenance and cue monitoring in younger and older adults' prospective memory}.
\newblock \bibinfo{journal}{\emph{Memory \& cognition}}  \bibinfo{volume}{45} (\bibinfo{date}{06} \bibinfo{year}{2017}).
\newblock
\urldef\tempurl%
\url{https://doi.org/10.3758/s13421-017-0720-5}
\showDOI{\tempurl}


\bibitem[Barner et~al\mbox{.}(2019)]%
        {Barner2019TimeofdayEO}
\bibfield{author}{\bibinfo{person}{Christine Barner}, \bibinfo{person}{Sarah~Rebecca Schmid}, {and} \bibinfo{person}{Susanne Diekelmann}.} \bibinfo{year}{2019}\natexlab{}.
\newblock \showarticletitle{Time-of-day effects on prospective memory}.
\newblock \bibinfo{journal}{\emph{Behavioural Brain Research}}  \bibinfo{volume}{376} (\bibinfo{year}{2019}).
\newblock
\urldef\tempurl%
\url{https://doi.org/10.1016/j.bbr.2019.112179}
\showDOI{\tempurl}


\bibitem[Bellotti et~al\mbox{.}(2004)]%
        {10.1145/985692.985785}
\bibfield{author}{\bibinfo{person}{Victoria Bellotti}, \bibinfo{person}{Brinda Dalal}, \bibinfo{person}{Nathaniel Good}, \bibinfo{person}{Peter Flynn}, \bibinfo{person}{Daniel~G. Bobrow}, {and} \bibinfo{person}{Nicolas Ducheneaut}.} \bibinfo{year}{2004}\natexlab{}.
\newblock \showarticletitle{What a To-Do: Studies of Task Management towards the Design of a Personal Task List Manager}. In \bibinfo{booktitle}{\emph{Proceedings of the SIGCHI Conference on Human Factors in Computing Systems}} (Vienna, Austria) \emph{(\bibinfo{series}{CHI '04})}. \bibinfo{publisher}{Association for Computing Machinery}, \bibinfo{address}{New York, NY, USA}, \bibinfo{pages}{735–742}.
\newblock
\showISBNx{1581137028}
\urldef\tempurl%
\url{https://doi.org/10.1145/985692.985785}
\showDOI{\tempurl}


\bibitem[Bellotti et~al\mbox{.}(2003)]%
        {bellotti_taskmaster}
\bibfield{author}{\bibinfo{person}{Victoria Bellotti}, \bibinfo{person}{Nicolas Ducheneaut}, \bibinfo{person}{Mark Howard}, {and} \bibinfo{person}{Ian Smith}.} \bibinfo{year}{2003}\natexlab{}.
\newblock \showarticletitle{Taking Email to Task: The Design and Evaluation of a Task Management Centered Email Tool}. In \bibinfo{booktitle}{\emph{Proceedings of the SIGCHI Conference on Human Factors in Computing Systems}} (Ft. Lauderdale, Florida, USA) \emph{(\bibinfo{series}{CHI '03})}. \bibinfo{publisher}{Association for Computing Machinery}, \bibinfo{address}{New York, NY, USA}, \bibinfo{pages}{345–352}.
\newblock
\showISBNx{1581136307}
\urldef\tempurl%
\url{https://doi.org/10.1145/642611.642672}
\showDOI{\tempurl}


\bibitem[Bellotti et~al\mbox{.}(2005)]%
        {bellotti2005quality}
\bibfield{author}{\bibinfo{person}{Victoria Bellotti}, \bibinfo{person}{Nicolas Ducheneaut}, \bibinfo{person}{Mark Howard}, \bibinfo{person}{Ian Smith}, {and} \bibinfo{person}{Rebecca~E Grinter}.} \bibinfo{year}{2005}\natexlab{}.
\newblock \showarticletitle{Quality versus quantity: E-mail-centric task management and its relation with overload}.
\newblock \bibinfo{journal}{\emph{Human--Computer Interaction}} \bibinfo{volume}{20}, \bibinfo{number}{1-2} (\bibinfo{year}{2005}), \bibinfo{pages}{89--138}.
\newblock
\urldef\tempurl%
\url{https://doi.org/10.1207/s15327051hci2001&2_4}
\showURL{%
\tempurl}


\bibitem[Brewer et~al\mbox{.}(2017)]%
        {brewer2017remember}
\bibfield{author}{\bibinfo{person}{Robin~N Brewer}, \bibinfo{person}{Meredith~Ringel Morris}, {and} \bibinfo{person}{Si{\^a}n~E Lindley}.} \bibinfo{year}{2017}\natexlab{}.
\newblock \showarticletitle{How to remember what to remember: exploring possibilities for digital reminder systems}.
\newblock \bibinfo{journal}{\emph{Proceedings of the ACM on interactive, mobile, wearable and ubiquitous technologies}} \bibinfo{volume}{1}, \bibinfo{number}{3} (\bibinfo{year}{2017}), \bibinfo{pages}{1--20}.
\newblock
\urldef\tempurl%
\url{https://doi.org/10.1145/3130903}
\showDOI{\tempurl}


\bibitem[Chen et~al\mbox{.}(2019)]%
        {chen2019effect}
\bibfield{author}{\bibinfo{person}{Jierong Chen}, \bibinfo{person}{Zhen Wei}, \bibinfo{person}{Hongying Han}, \bibinfo{person}{Lili Jin}, \bibinfo{person}{Chuanyong Xu}, \bibinfo{person}{Dan Dong}, \bibinfo{person}{Jianping Lu}, \bibinfo{person}{Guobin Wan}, {and} \bibinfo{person}{Ziwen Peng}.} \bibinfo{year}{2019}\natexlab{}.
\newblock \showarticletitle{An effect of chronic stress on prospective memory via alteration of resting-state hippocampal subregion functional connectivity}.
\newblock \bibinfo{journal}{\emph{Scientific Reports}} \bibinfo{volume}{9}, \bibinfo{number}{1} (\bibinfo{year}{2019}), \bibinfo{pages}{1--9}.
\newblock
\urldef\tempurl%
\url{https://doi.org/10.1038/s41598-019-56111-9}
\showDOI{\tempurl}


\bibitem[Corston-Oliver et~al\mbox{.}(2004)]%
        {corston2004task}
\bibfield{author}{\bibinfo{person}{Simon Corston-Oliver}, \bibinfo{person}{Eric Ringger}, \bibinfo{person}{Michael Gamon}, {and} \bibinfo{person}{Richard Campbell}.} \bibinfo{year}{2004}\natexlab{}.
\newblock \showarticletitle{Task-focused summarization of email}. In \bibinfo{booktitle}{\emph{Text Summarization Branches Out}}. \bibinfo{pages}{43--50}.
\newblock
\urldef\tempurl%
\url{https://aclanthology.org/W04-1008.pdf}
\showURL{%
\tempurl}


\bibitem[Dabbish and Kraut(2006)]%
        {dabbish2006email}
\bibfield{author}{\bibinfo{person}{Laura~A Dabbish} {and} \bibinfo{person}{Robert~E Kraut}.} \bibinfo{year}{2006}\natexlab{}.
\newblock \showarticletitle{Email overload at work: An analysis of factors associated with email strain}. In \bibinfo{booktitle}{\emph{Proceedings of the 2006 20th anniversary conference on Computer supported cooperative work}}. \bibinfo{pages}{431--440}.
\newblock
\urldef\tempurl%
\url{https://doi.org/10.1145/1180875.1180941}
\showDOI{\tempurl}


\bibitem[DeFilippis et~al\mbox{.}(2022)]%
        {DeFilippis_Impink_Singell_Polzer_Sadun_2022}
\bibfield{author}{\bibinfo{person}{Evan DeFilippis}, \bibinfo{person}{Stephen~Michael Impink}, \bibinfo{person}{Madison Singell}, \bibinfo{person}{Jeffrey~T. Polzer}, {and} \bibinfo{person}{Raffaella Sadun}.} \bibinfo{year}{2022}\natexlab{}.
\newblock \showarticletitle{The impact of COVID-19 on digital communication patterns}.
\newblock \bibinfo{journal}{\emph{Humanities and Social Sciences Communications}} \bibinfo{volume}{9}, \bibinfo{number}{1} (\bibinfo{date}{May} \bibinfo{year}{2022}), \bibinfo{pages}{1–11}.
\newblock
\showISSN{2662-9992}
\urldef\tempurl%
\url{https://doi.org/10.1057/s41599-022-01190-9}
\showDOI{\tempurl}


\bibitem[Dismukes(2012)]%
        {dismukes2012prospective}
\bibfield{author}{\bibinfo{person}{R~Key Dismukes}.} \bibinfo{year}{2012}\natexlab{}.
\newblock \showarticletitle{Prospective memory in workplace and everyday situations}.
\newblock \bibinfo{journal}{\emph{Current Directions in Psychological Science}} \bibinfo{volume}{21}, \bibinfo{number}{4} (\bibinfo{year}{2012}), \bibinfo{pages}{215--220}.
\newblock
\urldef\tempurl%
\url{https://doi.org/10.1177/0963721412447621}
\showDOI{\tempurl}


\bibitem[Ducheneaut and Bellotti(2001)]%
        {ducheneaut2001mail}
\bibfield{author}{\bibinfo{person}{Nicolas Ducheneaut} {and} \bibinfo{person}{Victoria Bellotti}.} \bibinfo{year}{2001}\natexlab{}.
\newblock \showarticletitle{E-mail as habitat: an exploration of embedded personal information management}.
\newblock \bibinfo{journal}{\emph{interactions}} \bibinfo{volume}{8}, \bibinfo{number}{5} (\bibinfo{year}{2001}), \bibinfo{pages}{30--38}.
\newblock
\urldef\tempurl%
\url{https://doi.org/10.1145/382899.383305}
\showDOI{\tempurl}


\bibitem[Einstein and McDaniel(1990)]%
        {Einstein1990NormalAA}
\bibfield{author}{\bibinfo{person}{Gilles~O. Einstein} {and} \bibinfo{person}{Mark~A. McDaniel}.} \bibinfo{year}{1990}\natexlab{}.
\newblock \showarticletitle{Normal aging and prospective memory.}
\newblock \bibinfo{journal}{\emph{Journal of experimental psychology. Learning, memory, and cognition}}  \bibinfo{volume}{16 4} (\bibinfo{year}{1990}), \bibinfo{pages}{717--26}.
\newblock
\urldef\tempurl%
\url{https://doi.org/10.1037/0278-7393.16.4.717}
\showDOI{\tempurl}


\bibitem[Einstein et~al\mbox{.}(1997)]%
        {einstein1997aging}
\bibfield{author}{\bibinfo{person}{Gilles~O Einstein}, \bibinfo{person}{Rebekah~E Smith}, \bibinfo{person}{Mark~A McDaniel}, {and} \bibinfo{person}{Pat Shaw}.} \bibinfo{year}{1997}\natexlab{}.
\newblock \showarticletitle{Aging and prospective memory: the influence of increased task demands at encoding and retrieval.}
\newblock \bibinfo{journal}{\emph{Psychology and aging}} \bibinfo{volume}{12}, \bibinfo{number}{3} (\bibinfo{year}{1997}), \bibinfo{pages}{479}.
\newblock
\urldef\tempurl%
\url{https://doi.org/10.1037/0882-7974.12.3.479}
\showDOI{\tempurl}


\bibitem[Eldridge et~al\mbox{.}(1992)]%
        {eldridge1992memory}
\bibfield{author}{\bibinfo{person}{Margery Eldridge}, \bibinfo{person}{Abigail Sellen}, {and} \bibinfo{person}{Debra Bekerian}.} \bibinfo{year}{1992}\natexlab{}.
\newblock \showarticletitle{Memory problems at work: Their range, frequency and severity}.
\newblock \bibinfo{journal}{\emph{Rank Xerox, EuroPARC}} (\bibinfo{year}{1992}).
\newblock
\urldef\tempurl%
\url{https://www.microsoft.com/en-us/research/wp-content/uploads/2016/08/memory-problems-92.pdf}
\showURL{%
\tempurl}


\bibitem[Faulring et~al\mbox{.}(2010)]%
        {faulring2010agent}
\bibfield{author}{\bibinfo{person}{Andrew Faulring}, \bibinfo{person}{Brad Myers}, \bibinfo{person}{Ken Mohnkern}, \bibinfo{person}{Bradley Schmerl}, \bibinfo{person}{Aaron Steinfeld}, \bibinfo{person}{John Zimmerman}, \bibinfo{person}{Asim Smailagic}, \bibinfo{person}{Jeffery Hansen}, {and} \bibinfo{person}{Daniel Siewiorek}.} \bibinfo{year}{2010}\natexlab{}.
\newblock \showarticletitle{Agent-assisted task management that reduces email overload}. In \bibinfo{booktitle}{\emph{Proceedings of the 15th international conference on Intelligent user interfaces}}. \bibinfo{pages}{61--70}.
\newblock
\urldef\tempurl%
\url{https://doi.org/10.1145/1719970.1719980}
\showDOI{\tempurl}


\bibitem[Gilbert(2015)]%
        {gilbert2015strategic}
\bibfield{author}{\bibinfo{person}{Sam~J Gilbert}.} \bibinfo{year}{2015}\natexlab{}.
\newblock \showarticletitle{Strategic offloading of delayed intentions into the external environment}.
\newblock \bibinfo{journal}{\emph{Quarterly journal of experimental psychology}} \bibinfo{volume}{68}, \bibinfo{number}{5} (\bibinfo{year}{2015}), \bibinfo{pages}{971--992}.
\newblock


\bibitem[Gilbert et~al\mbox{.}(2022)]%
        {Gilbert2022OutsourcingMT}
\bibfield{author}{\bibinfo{person}{Sam~J. Gilbert}, \bibinfo{person}{Annika Boldt}, \bibinfo{person}{Chhavi Sachdeva}, \bibinfo{person}{Chiara Scarampi}, {and} \bibinfo{person}{Pei-Chun Tsai}.} \bibinfo{year}{2022}\natexlab{}.
\newblock \showarticletitle{Outsourcing Memory to External Tools: A Review of 'Intention Offloading'.}
\newblock \bibinfo{journal}{\emph{Psychonomic bulletin \& review}} (\bibinfo{year}{2022}).
\newblock
\urldef\tempurl%
\url{https://doi.org/10.31234/osf.io/ahqtz}
\showDOI{\tempurl}


\bibitem[Haas et~al\mbox{.}(2020)]%
        {haas2020}
\bibfield{author}{\bibinfo{person}{Maximilian Haas}, \bibinfo{person}{Sascha Zuber}, \bibinfo{person}{Matthias Kliegel}, {and} \bibinfo{person}{Nicola Ballhausen}.} \bibinfo{year}{2020}\natexlab{}.
\newblock \showarticletitle{Prospective memory errors in everyday life: Does instruction matter?}
\newblock \bibinfo{journal}{\emph{Memory}} \bibinfo{volume}{28}, \bibinfo{number}{2} (\bibinfo{year}{2020}), \bibinfo{pages}{196--203}.
\newblock
\showISSN{0965-8211}
\urldef\tempurl%
\url{https://doi.org/10.1080/09658211.2019.1707227}
\showDOI{\tempurl}


\bibitem[Horvitz(2022)]%
        {horvitz2022horizon}
\bibfield{author}{\bibinfo{person}{Eric Horvitz}.} \bibinfo{year}{2022}\natexlab{}.
\newblock \showarticletitle{On the horizon: {I}nteractive and compositional deepfakes}. In \bibinfo{booktitle}{\emph{Proceedings of the 2022 International Conference on Multimodal Interaction}}. \bibinfo{pages}{653--661}.
\newblock
\urldef\tempurl%
\url{https://dl.acm.org/doi/abs/10.1145/3536221.3558175}
\showURL{%
\tempurl}


\bibitem[Horvitz et~al\mbox{.}(2004)]%
        {horvitz2004learning}
\bibfield{author}{\bibinfo{person}{Eric Horvitz}, \bibinfo{person}{Susan Dumais}, {and} \bibinfo{person}{Paul Koch}.} \bibinfo{year}{2004}\natexlab{}.
\newblock \showarticletitle{Learning predictive models of memory landmarks}. In \bibinfo{booktitle}{\emph{Proceedings of the Annual Meeting of the Cognitive Science Society}}, Vol.~\bibinfo{volume}{26}.
\newblock
\urldef\tempurl%
\url{https://doi.org/10.19066/cogsci.2009.20.4.006}
\showDOI{\tempurl}


\bibitem[Horvitz et~al\mbox{.}(1999)]%
        {Priorities1999}
\bibfield{author}{\bibinfo{person}{Eric Horvitz}, \bibinfo{person}{Andy Jacobs}, {and} \bibinfo{person}{David Hovel}.} \bibinfo{year}{1999}\natexlab{}.
\newblock \showarticletitle{Attention-Sensitive Alerting}. In \bibinfo{booktitle}{\emph{Proceedings of the Fifteenth Conference on Uncertainty in Artificial Intelligence}} (Stockholm, Sweden). \bibinfo{pages}{305–313}.
\newblock


\bibitem[Horvitz(2001)]%
        {horvitz2001notification}
\bibfield{author}{\bibinfo{person}{Edward Cutrell Mary Czerwinski~Eric Horvitz}.} \bibinfo{year}{2001}\natexlab{}.
\newblock \showarticletitle{Notification, disruption, and memory: Effects of messaging interruptions on memory and performance}. In \bibinfo{booktitle}{\emph{Human-Computer Interaction: INTERACT}}, Vol.~\bibinfo{volume}{1}. \bibinfo{pages}{263}.
\newblock
\urldef\tempurl%
\url{https://doi.org/ftp/Interact2001.pdf}
\showDOI{\tempurl}


\bibitem[Ihle et~al\mbox{.}(2012)]%
        {ihle2012age}
\bibfield{author}{\bibinfo{person}{Andreas Ihle}, \bibinfo{person}{Katharina Schnitzspahn}, \bibinfo{person}{Peter~G Rendell}, \bibinfo{person}{C{\"a}cilia Luong}, {and} \bibinfo{person}{Matthias Kliegel}.} \bibinfo{year}{2012}\natexlab{}.
\newblock \showarticletitle{Age benefits in everyday prospective memory: The influence of personal task importance, use of reminders and everyday stress}.
\newblock \bibinfo{journal}{\emph{Aging, Neuropsychology, and Cognition}} \bibinfo{volume}{19}, \bibinfo{number}{1-2} (\bibinfo{year}{2012}), \bibinfo{pages}{84--101}.
\newblock
\urldef\tempurl%
\url{https://doi.org/10.1080/13825585.2011.629288}
\showDOI{\tempurl}


\bibitem[Iqbal and Horvitz(2007)]%
        {iqbal2007disruption}
\bibfield{author}{\bibinfo{person}{Shamsi~T Iqbal} {and} \bibinfo{person}{Eric Horvitz}.} \bibinfo{year}{2007}\natexlab{}.
\newblock \showarticletitle{Disruption and recovery of computing tasks: field study, analysis, and directions}. In \bibinfo{booktitle}{\emph{Proceedings of the SIGCHI conference on Human factors in computing systems}}. \bibinfo{pages}{677--686}.
\newblock
\urldef\tempurl%
\url{https://doi.org/10.1145/1240624.1240730}
\showDOI{\tempurl}


\bibitem[Iqbal and Horvitz(2010)]%
        {iqbal2010notifications}
\bibfield{author}{\bibinfo{person}{Shamsi~T Iqbal} {and} \bibinfo{person}{Eric Horvitz}.} \bibinfo{year}{2010}\natexlab{}.
\newblock \showarticletitle{Notifications and awareness: a field study of alert usage and preferences}. In \bibinfo{booktitle}{\emph{Proceedings of the 2010 ACM conference on Computer supported cooperative work}}. \bibinfo{pages}{27--30}.
\newblock
\urldef\tempurl%
\url{https://doi.org/10.1145/1718918.1718926}
\showDOI{\tempurl}


\bibitem[Khaokaew et~al\mbox{.}(2022)]%
        {khaokaew2022imagining}
\bibfield{author}{\bibinfo{person}{Yonchanok Khaokaew}, \bibinfo{person}{Indigo Holcombe-James}, \bibinfo{person}{Mohammad~Saiedur Rahaman}, \bibinfo{person}{Jonathan Liono}, \bibinfo{person}{Johanne~R. Trippas}, \bibinfo{person}{Damiano Spina}, \bibinfo{person}{Peter Bailey}, \bibinfo{person}{Nicholas~J. Belkin}, \bibinfo{person}{Paul~N. Bennett}, \bibinfo{person}{Yongli Ren}, \bibinfo{person}{Mark Sanderson}, \bibinfo{person}{Falk Scholer}, \bibinfo{person}{Ryen~W. White}, {and} \bibinfo{person}{Flora~D. Salim}.} \bibinfo{year}{2022}\natexlab{}.
\newblock \showarticletitle{Imagining Future Digital Assistants at Work: A Study of Task Management Needs}.
\newblock \bibinfo{journal}{\emph{International Journal of Human-Computer Studies}} (\bibinfo{year}{2022}).
\newblock
\urldef\tempurl%
\url{https://doi.org/10.1016/j.ijhcs.2022.102905}
\showDOI{\tempurl}


\bibitem[Kimani et~al\mbox{.}(2019)]%
        {kimani2019conversational}
\bibfield{author}{\bibinfo{person}{Everlyne Kimani}, \bibinfo{person}{Kael Rowan}, \bibinfo{person}{Daniel McDuff}, \bibinfo{person}{Mary Czerwinski}, {and} \bibinfo{person}{Gloria Mark}.} \bibinfo{year}{2019}\natexlab{}.
\newblock \showarticletitle{A conversational agent in support of productivity and wellbeing at work}. In \bibinfo{booktitle}{\emph{2019 8th international conference on affective computing and intelligent interaction (ACII)}}. IEEE, \bibinfo{pages}{1--7}.
\newblock
\urldef\tempurl%
\url{https://doi.org/10.1109/acii.2019.8925488}
\showDOI{\tempurl}


\bibitem[Kokkalis et~al\mbox{.}(2013)]%
        {kokkalis2013emailvalet}
\bibfield{author}{\bibinfo{person}{Nicolas Kokkalis}, \bibinfo{person}{Thomas K{\"o}hn}, \bibinfo{person}{Carl Pfeiffer}, \bibinfo{person}{Dima Chornyi}, \bibinfo{person}{Michael~S Bernstein}, {and} \bibinfo{person}{Scott~R Klemmer}.} \bibinfo{year}{2013}\natexlab{}.
\newblock \showarticletitle{EmailValet: Managing email overload through private, accountable crowdsourcing}. In \bibinfo{booktitle}{\emph{Proceedings of the 2013 conference on Computer supported cooperative work}}. \bibinfo{pages}{1291--1300}.
\newblock
\urldef\tempurl%
\url{https://doi.org/10.1145/2441776.2441922}
\showDOI{\tempurl}


\bibitem[Kr{\"a}mer(2010)]%
        {kramer2010pim}
\bibfield{author}{\bibinfo{person}{Jan-Peter Kr{\"a}mer}.} \bibinfo{year}{2010}\natexlab{}.
\newblock \showarticletitle{PIM-Mail: consolidating task and email management}.
\newblock In \bibinfo{booktitle}{\emph{CHI'10 Extended Abstracts on Human Factors in Computing Systems}}. \bibinfo{pages}{4411--4416}.
\newblock
\urldef\tempurl%
\url{https://doi.org/10.1145/1753846.1754162}
\showDOI{\tempurl}


\bibitem[Mackay(1988)]%
        {mackay1988more}
\bibfield{author}{\bibinfo{person}{Wendy~E Mackay}.} \bibinfo{year}{1988}\natexlab{}.
\newblock \showarticletitle{More than just a communication system: diversity in the use of electronic mail}. In \bibinfo{booktitle}{\emph{Proceedings of the 1988 ACM conference on Computer-supported cooperative work}}. \bibinfo{pages}{344--353}.
\newblock
\urldef\tempurl%
\url{https://doi.org/10.1145/62266.62293}
\showURL{%
\tempurl}


\bibitem[Manseau(2020)]%
        {manseau2020ai}
\bibfield{author}{\bibinfo{person}{Jasmin Manseau}.} \bibinfo{year}{2020}\natexlab{}.
\newblock \showarticletitle{AI in the workplace: A qualitative analysis of intelligent employee assistants}.
\newblock  (\bibinfo{year}{2020}).
\newblock
\urldef\tempurl%
\url{https://core.ac.uk/download/pdf/326836018.pdf}
\showURL{%
\tempurl}


\bibitem[McMurtry(2014)]%
        {mcmurtry2014managing}
\bibfield{author}{\bibinfo{person}{Kim McMurtry}.} \bibinfo{year}{2014}\natexlab{}.
\newblock \showarticletitle{Managing email overload in the workplace}.
\newblock \bibinfo{journal}{\emph{Performance Improvement}} \bibinfo{volume}{53}, \bibinfo{number}{7} (\bibinfo{year}{2014}), \bibinfo{pages}{31--37}.
\newblock
\urldef\tempurl%
\url{https://doi.org/10.1002/pfi.21424}
\showDOI{\tempurl}


\bibitem[Meyerhoff et~al\mbox{.}(2021)]%
        {meyerhoff2021individual}
\bibfield{author}{\bibinfo{person}{Hauke~S Meyerhoff}, \bibinfo{person}{Sandra Grinschgl}, \bibinfo{person}{Frank Papenmeier}, {and} \bibinfo{person}{Sam~J Gilbert}.} \bibinfo{year}{2021}\natexlab{}.
\newblock \showarticletitle{Individual differences in cognitive offloading: A comparison of intention offloading, pattern copy, and short-term memory capacity}.
\newblock \bibinfo{journal}{\emph{Cognitive Research: Principles and Implications}} \bibinfo{volume}{6}, \bibinfo{number}{1} (\bibinfo{year}{2021}), \bibinfo{pages}{34}.
\newblock


\bibitem[Mukherjee et~al\mbox{.}(2020)]%
        {mukherjee2020smart}
\bibfield{author}{\bibinfo{person}{Sudipto Mukherjee}, \bibinfo{person}{Subhabrata Mukherjee}, \bibinfo{person}{Marcello Hasegawa}, \bibinfo{person}{Ahmed~Hassan Awadallah}, {and} \bibinfo{person}{Ryen White}.} \bibinfo{year}{2020}\natexlab{}.
\newblock \showarticletitle{Smart to-do: Automatic generation of to-do items from emails}.
\newblock \bibinfo{journal}{\emph{arXiv preprint arXiv:2005.06282}} (\bibinfo{year}{2020}).
\newblock


\bibitem[Paul et~al\mbox{.}(2015)]%
        {paul2015interruptive}
\bibfield{author}{\bibinfo{person}{Celeste~Lyn Paul}, \bibinfo{person}{Anita Komlodi}, {and} \bibinfo{person}{Wayne Lutters}.} \bibinfo{year}{2015}\natexlab{}.
\newblock \showarticletitle{Interruptive notifications in support of task management}.
\newblock \bibinfo{journal}{\emph{International Journal of Human-Computer Studies}}  \bibinfo{volume}{79} (\bibinfo{year}{2015}), \bibinfo{pages}{20--34}.
\newblock
\urldef\tempurl%
\url{https://doi.org/10.1016/j.ijhcs.2015.02.001}
\showDOI{\tempurl}


\bibitem[Shimamura et~al\mbox{.}(1991)]%
        {Shimamura1991WhatIT}
\bibfield{author}{\bibinfo{person}{Arthur~P. Shimamura}, \bibinfo{person}{Jeri~S. Janowsky}, {and} \bibinfo{person}{Larry~R. Squire}.} \bibinfo{year}{1991}\natexlab{}.
\newblock \showarticletitle{What is the role of frontal lobe damage in memory disorders}.
\newblock


\bibitem[Toxtli et~al\mbox{.}(2018)]%
        {10.1145/3173574.3173632}
\bibfield{author}{\bibinfo{person}{Carlos Toxtli}, \bibinfo{person}{Andr\'{e}s Monroy-Hern\'{a}ndez}, {and} \bibinfo{person}{Justin Cranshaw}.} \bibinfo{year}{2018}\natexlab{}.
\newblock \showarticletitle{Understanding Chatbot-Mediated Task Management}. In \bibinfo{booktitle}{\emph{Proceedings of the 2018 CHI Conference on Human Factors in Computing Systems}} (Montreal QC, Canada) \emph{(\bibinfo{series}{CHI '18})}. \bibinfo{publisher}{Association for Computing Machinery}, \bibinfo{address}{New York, NY, USA}, \bibinfo{pages}{1–6}.
\newblock
\showISBNx{9781450356206}
\urldef\tempurl%
\url{https://doi.org/10.1145/3173574.3173632}
\showDOI{\tempurl}


\bibitem[Trippas et~al\mbox{.}(2019)]%
        {trippas2019learning}
\bibfield{author}{\bibinfo{person}{Johanne~R Trippas}, \bibinfo{person}{Damiano Spina}, \bibinfo{person}{Falk Scholer}, \bibinfo{person}{Ahmed~Hassan Awadallah}, \bibinfo{person}{Peter Bailey}, \bibinfo{person}{Paul~N Bennett}, \bibinfo{person}{Ryen~W White}, \bibinfo{person}{Jonathan Liono}, \bibinfo{person}{Yongli Ren}, \bibinfo{person}{Flora~D Salim}, {et~al\mbox{.}}} \bibinfo{year}{2019}\natexlab{}.
\newblock \showarticletitle{Learning about work tasks to inform intelligent assistant design}. In \bibinfo{booktitle}{\emph{Proceedings of the 2019 Conference on Human Information Interaction and Retrieval}}. \bibinfo{pages}{5--14}.
\newblock
\urldef\tempurl%
\url{https://doi.org/10.1145/3295750.3298934}
\showDOI{\tempurl}


\bibitem[Walter and Meier(2014)]%
        {10.3389/fpsyg.2014.00657}
\bibfield{author}{\bibinfo{person}{Stefan Walter} {and} \bibinfo{person}{Beat Meier}.} \bibinfo{year}{2014}\natexlab{}.
\newblock \showarticletitle{How important is importance for prospective memory? A review}.
\newblock \bibinfo{journal}{\emph{Frontiers in Psychology}}  \bibinfo{volume}{5} (\bibinfo{year}{2014}).
\newblock
\showISSN{1664-1078}
\urldef\tempurl%
\url{https://doi.org/10.3389/fpsyg.2014.00657}
\showDOI{\tempurl}


\bibitem[Whittaker and Sidner(1996)]%
        {whittaker1996email}
\bibfield{author}{\bibinfo{person}{Steve Whittaker} {and} \bibinfo{person}{Candace Sidner}.} \bibinfo{year}{1996}\natexlab{}.
\newblock \showarticletitle{Email overload: exploring personal information management of email}. In \bibinfo{booktitle}{\emph{Proceedings of the SIGCHI conference on Human factors in computing systems}}. \bibinfo{pages}{276--283}.
\newblock
\urldef\tempurl%
\url{https://doi.org/10.4324/9781315806389-24}
\showDOI{\tempurl}


\end{thebibliography}
